# PMm$^2$: large photomultipliers and innovative electronics for the next-generation neutrino experiments

B. Genolini[b,*], P. Barrillon[a], S. Blin[a], J.-E. Campagne[a], B. Combettes[d], S. Conforti[a], A.-G. Dehaine[d], D. Duchesneau[c], F. Dulucq[a], N. Dumont-Dayot[c], J. Favier[c], F. Fouché[d], R. Hermel[c], C. de La Taille[a], G. Martin-Chassard[a], T. Nguyen Trung[b], C. Périnet[b], J. Peyré[b], J. Pouthas[b], L. Raux[a], E. Rindel[b], P. Rosier[b], J. Tassan-Viol[c], W. Wei[a], A. Zghiche[c]

*Corresponding author: Bernard Genolini, genolini@ipno.in2p3.fr - Phone: +33 1 69 15 63 88 - Fax: +33 1 69 15 50 01

[a]Laboratoire de l'Accélérateur Linéaire, IN2P3-CNRS, Université Paris-Sud, Bat. 200, 91898 Orsay Cedex, France
[b]Institut de Physique Nucléaire d'Orsay, IN2P3-CNRS, Université Paris-Sud, 91406 Orsay Cedex, France
[c]LAPP Annecy, IN2P3-CNRS, Université de Haute Savoie, 9 chemin de Bellevue, BP 110, 74941 Annecy-le-Vieux Cedex, France
[d]Photonis France SAS, Avenue Roger Roncier, BP 520, 19100 Brive La Gaillarde, France

**Abstract**

The next generation of proton decay and neutrino experiments, the post-SuperKamiokande detectors as those that will take place in megaton size water tanks, will require very large surfaces of photodetection and a large volume of data. Even with large hemispherical photomultiplier tubes, the expected number of channels should reach hundreds of thousands. A funded R&D program to implement a solution is presented here. The very large surface of photodetection is segmented in macro pixels made of 16 hemispherical (12 inches) photomultiplier tubes connected to an autonomous front-end which works on a triggerless data acquisition mode. The expected data transmission rate is 5 Mb/s per cable, which can be achieved with existing techniques. This architecture allows to reduce considerably the cost and facilitate the industrialization.

This document presents the simulations and measurements which define the requirements for the photomultipliers and the electronics. A prototype of front-end electronics was successfully tested with 16 photomultiplier tubes supplied by a single high voltage, validating the built-in gain adjustment and the calibration principle. The first tests and calculations on the photomultiplier glass led to the study of a new package optimized for a 10 bar pressure in order to sustain the high underwater pressure.

Classification codes: 30.000 High Energy Physics Systems, 40.000 Space Rad. and Neutrino Detectors, 80.000 Data acquisition and Control, 90.000 Electronics

## 1. Introduction

The next generation of neutrino experiments [1] will be extensions by 10 to 20 times of the Super-Kamiokande detector [2]. The photodetection will rely on hundreds of thousands of hemispherical photomultiplier tubes (PMTs) covering the surface of the tanks. As a consequence, the design has to be driven by the costs and integrate industrial assembly procedures. A cost and performance analysis [3] showed that important savings could be obtained by using 12-inch diameter PMTs instead of the 20-inch ones used in Super-Kamiokande. However, the increase in the number of electronics channel has to be compensated by a change in the electronics and data acquisition architecture.

This document presents a R&D project [4] which proposes such an architecture based on a triggerless data acquisition and an innovative front-end electronics. The project name, PMm$^2$, summarizes the idea of paving the detector surface with square meter (or so) photodetection units. This R&D is carried out by three laboratories (LAL Orsay, IPN Orsay, LAPP Annecy) and the Photonis company [5], the French photomultiplier tube manufacturer. It is funded by the French National Agency for Research (ANR) under the reference ANR-06-BLAN-0186.

## 2. The system architecture

The next generation of proton decay and neutrino experiments will comprise hundred of thousands of photomultiplier tubes (PMTs) in large Cerenkov water tanks [1]. For instance, the MEMPHYS project aims at constructing three or four Cerenkov water tanks, comprising about 80 000 PMTs each [6]. The expected water height may be as important as 65 m, which creates a maximum pressure of 7.5 bar. Therefore, the requirement in the PMm$^2$ project was set to 10 bars for safety.

### 2.1 Cost considerations

The experiment will use hemispherical PMTs which are the largest existing photodetectors. The PMm$^2$ project proposes to use 12-inch diameter tubes, instead of the 20 inch ones as in SuperKamiokande [2]. Though their photodetection surface is smaller, the 12-inch PMTs have a greater photodetection efficiency, a better reliability and they are easier to produce. It has been estimated that the cost per surface area and detected photoelectron is reduced by a factor of 1.6 by using PMTs with a diameter of 12 inches instead of 20 inches [3].

However, the cost reduction due to the smaller PMT has to face the increase in the number of electronics channels. Therefore, an additional saving is proposed by grouping the PMTs by 16, as shown in Figure 1. They will share a common electronics, a common high voltage bias distributed by a common front-end module located close to the tubes. A data cable for the connection with the surface will also carry the power voltage. The front-end electronics will consist in an ASIC (Application Specific Integrated Circuit) coupled to a FPGA (Field-Programmable Gate Array) that will manage the dialog with the surface controller through the data cable. At this scale of production, costs will be highly reduced by designing the ASIC as a System on a Chip (SoC), i.e. processing the analog signal up to the digitization. This SoC achieves also the compensation of the PMT gain dispersion due to the common high voltage.

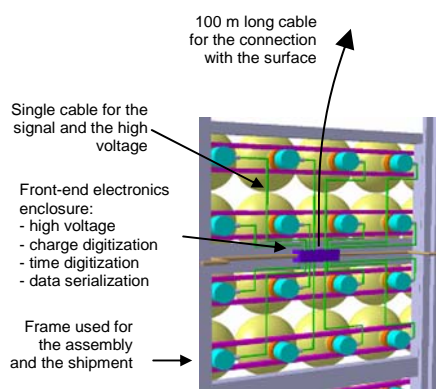

Figure 1 Principle of the PMm$^2$ architecture: the PMTs are gathered by groups of 16, sharing a common front-end electronics and connection cable to the surface.

The 16 PMTs will be assembled on a frame which will simplify the production, the shipment and the integration to the detector. The connectors between the PMTs and the front-end module can be suppressed, thus leading to an important additional saving, because the underwater connectors are expensive. The only connector per group of 16 PMTs would be for the connection to the cable that links the front-end module with the surface.

## 2.2 A triggerless data acquisition

The detection relies on a Cerenkov cone that will produce on the surface of the tank in most of the cases a large circle much larger than a 16 PMT cell. Therefore it is impossible to provide a local trigger to the 16 channel electronics module so that it was decided to get the charge and the arrival time of all the pulses of each PMT. This triggerless mode implies a time tagging in order to perform time coincidence and to obtain a good resolution for the reconstruction of the direction of the incident particle. The requirement on the time resolution for the electronics is 1 ns since the expected single electron jitter is not better than 3 ns for the considered size of PMT. This performance is achieved by transmitting a 10 MHz clock synchronized by GPS to the front-end ASIC which counts the clock cycles and then measures the time between the clock pulses and the PMT pulses with a TDC. Since the front-end is located underwater, close to the PMTs, the 10 MHz clock is transmitted through the 100 m long data cable that connects the front-end module to the surface. A time calibration is necessary to measure the transit time of each cable, and it could be performed with a global calibration similar to that of the Super-Kamiokande experiment [2].

## 3. An underwater front-end

The front-end electronics will be in an enclosure located underwater close to the PMTs. It has to achieve the PMT bias, the slow control, the data digitization and the interface with the cable connection to the surface.

The power (48 V with a maximum of 20 W) is provided by the data cable. A single high voltage converter produces the high voltage (at maximum 2 kV) to bias the PMTs. The same coaxial cable is used for the high voltage and the signals. Measurements performed with a 2 meter long cable showed that this solution adds an independent noise less than $10^4$ electrons, thus negligible at a working gain greater than $10^6$.

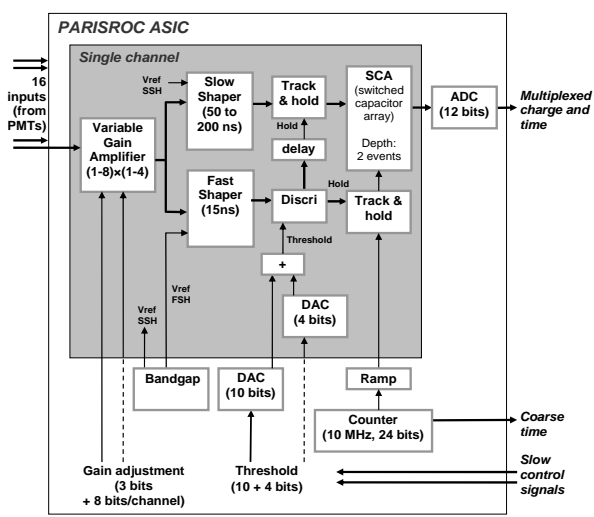

Figure 2 Architecture of the PARISROC chip. It is composed of 16 independent channels featuring the same functionnalities (gray zone on the figure). The functionnalities implemented for test purposes are not shown on this figure.

The front-end electronics consists of an ASIC called PARISROC, produced with the 0.35 µm SiGe AMS technology. A first version, of which the surface is 5×3.4 mm$^2$, was submitted in June 2008. The architecture is presented in Figure 2. It achieves the compensation of the PMT gain dispersion, the charge digitization and the event time tagging for 16 independent channels. In this first version, the digitization is carried out by a 12 bit ADC shared by the 16 channels. All the output data are read out in serial mode. It is a self-triggered device: the discriminator used for the time tagging triggers the charge digitization. The main challenge for this version is to demonstrate the feasibility of 16 independent channels working in a SoC architecture.

### 3.1 Principle of the time and charge measurement

The event time tagging is achieved by triggering the signal after a fast shaper with a time constant of 15 ns. The threshold is programmable in order to trigger the PMT signal at a level of 0.3 photoelectrons. The time tag is calculated by counting over 24 bits the pulses from an external 10 MHz clock (the coarse time measurement), and by measuring over 12 bits the time difference between the latest clock pulse and the signal occurrence, with the technique implemented in the SPIROC ASIC [7]. This fine time difference is measured by holding a ramp signal shared by the 16 channels, which is triggered by the 10 MHz clock. This value is then stored in a switched capacitor array (SCA).

For the charge measurement, the signal is preamplified with an adjustable gain. It is then processed by a $CR-RC^2$ shaper with an adjustable time constant (50, 100 or 200 ns). The signal is held by the delayed trigger output, and stored in the SCA. The principle of the preamplifier, track and hold, and trigger architecture are those implemented on the MAROC2 ASIC [8]. They have been tested with this circuit and 10-inch PMTs (Photonis XP1804) by acquiring the single electron responses independently with a MAROC2 board and a LeCroy 2249A ADC CAMAC module. The comparison is presented in Figure 3 and shows a good agreement.

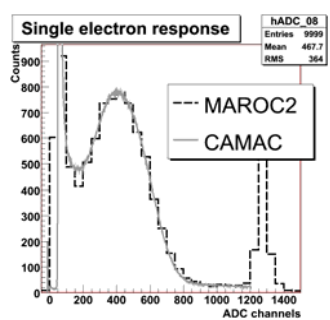

Figure 3 Comparison of the acquisition of a 10-inch PMT single electron response with a standard CAMAC acquisition (LeCroy 2249A ADC board) and with a MAROC2 board.

The values stored in the SCAs are multiplexed and digitized by a common Wilkinson ADC, which is adapted from that of the SPIROC ASIC [7] with some updates. It comprises a common 103 µs ramp and a common counter for the time and the charge. The 16 charges and 16 times are converted in a single ADC run. Its maximum conversion time is 103 µs. The total linearity (including the shaping, the SCA and the digitization) is better than 1 %.

### 3.2 Dead time and occurrence randomness compensation

The dead time is a critical issue of a triggerless data acquisition. With large PMTs, the noise rate at a level of 0.3 photoelectron is large and constitutes the main cause of dead time. The noise pulses occur independently on each channel following a Poisson process with a mean rate of a few kHz. The current ADC technology compatible with an implementation in a SoC generates a dead time with an order of magnitude of 100 µs, so that the digitization of the noise would reduce drastically the PMT detection efficiency. Therefore, it was decided to implement on each independent channel switched capacitor arrays (SCA) that will be used as FIFO buffers. The results obtained on the first ASIC prototypes will drive the dead time optimization (SCA depth, number of ADCs).

### 3.3 PMT gain compensation

A system compensating the different gains of the 16 PMTs has been implemented in the ASIC. It was estimated from a study from the production of Photonis tubes that the gain dispersion of tubes at a given voltage is such that the ratio between the highest and the lowest gain is not more than 12. After ageing, this ratio is multiplied by a factor of two. It is possible for the manufacturer to sort the PMTs at a reasonable cost when they are produced at a very large scale: the gain ratio can be reduced to 6 in a batch of 16 PMTs. The gain of the front-end electronics first stage can be set from 1 to 8 by 3 bits, common to all channels. The relative gain for each channel is currently adjustable from 1 to 4, with an accuracy of 8 bits. The threshold is set by a common 10 bit DAC, to which is added a voltage set by a 4 bit DAC specific to each channel. This compensation was simulated from the production data of 1 150 photomultiplier tubes. The relative error on the single photoelectron discrimination was estimated to be less than 1 % RMS.

## 4. R&D on PMT and data transmission

### 4.1 A new 12-inch PMT

The specifications for the PMTs are governed by the physics and the immersion in deep water. Simulations on the dynamic range and on the electronics calibration methods showed that their gain have to be set to a few $10^6$. Mechanical calculations were performed on the existing 12-inch tubes produced by Photonis, showing that they do not sustain the required 10 bar pressure. Therefore, Photonis changed the envelope design in order to sustain the pressure with the smallest glass thickness in order to limit the PMT noise due to the glass radioactivity. As a consequence the multiplier design had also to be changed because simulations showed that the collection was affected by the change of shape.

The PMT-base water tightness is ensured by a potting inside a waterproof enclosure. The tubes and their bases will be tested in a pressure vessel at a pressure up to 10 bars. This vessel is loaned to IPN Orsay by the Brookhaven National Laboratory. It will also be used to test the front-end enclosure and cable tightness.

### 4.2 The data transmission

The data transmission from the front-end module to the surface is carried out by a 100-m long cable, which serves also to manage the slow control, and to transmit the 10 MHz reference clock, power to supply the front-end module. The digitized data are transmitted with a dedicated serial protocol at a rate of 5 Mbit/s that was determined assuming that the noise for a single PMT is 5 kHz, and that there are 52 bits of data per PMT. Such a cable is made of at least two twisted pairs. One of them was tested successfully at up to 30 Mbit/s.

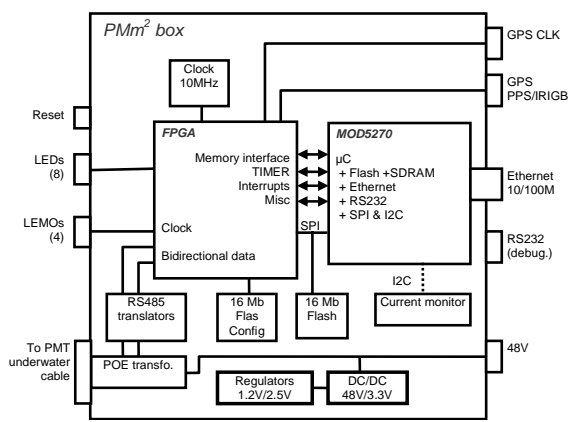

Figure 4 Architecture of the surface controller.

The data will be processed at the surface level on a network, of which the interface with the front-end is carried out by a surface controller. Its architecture is presented in Figure 4. It is a board that receives the data from one data cable using a dedicated serial protocol. It transmits also the 48 V sent to the front-end from an external power supply. It manages the 10 MHz clock synchronization from a GPS signal. The dialog with the main system is achieved in TCP/IP, through a 10/100M Ethernet interface.

## 5. Conclusion

PMm$^2$, a new concept for the next generation of megaton scale detectors foreseen for proton decay search and neutrino physics, was described in this document. The main innovative features are the reduction of the PMT diameter to 12 inches, a modular design (assembly by 16 PMTs), an underwater front-end electronics and a triggerless data acquisition. Most of the critical items have been validated and a first version of the front-end ASIC has been submitted. The water tightness at 10 bar will be tested with a dedicated pressure vessel.

The last phase of this project is the assembly of a demonstrator. A module of 16 12-inch PMTs, equipped with a front-end module, connected with a 100-meter long cable to a surface controller will be tested in a Cerenkov water tank on cosmic rays. All the functionalities of the projects will be tested, including the front-end ASIC, the power over data cable, and the clock synchronization with GPS. The validation experiment is foreseen by the end of 2009.